\documentclass[twocolumn,twoside,slac_two]{revtex4}
\usepackage{graphicx}
\usepackage{fancyhdr}
\usepackage[T1]{fontenc}
\usepackage[latin9]{inputenc}
\usepackage{ae,aecompl}
\providecommand{\tabularnewline}{\\}

\def\gsim{\mathrel{\raise.5ex\hbox{$>$}\mkern-14mu
             \lower0.6ex\hbox{$\sim$}}}
\def\lsim{\mathrel{\raise.3ex\hbox{$<$}\mkern-14mu
             \lower0.6ex\hbox{$\sim$}}}

\begin{document}

\title{Constraining Lorentz Invariance Violation with \textit{Fermi}}

\author{Vlasios Vasileiou for the \textit{Fermi} LAT and GBM Collaborations}
\affiliation{NASA Goddard Space Flight Center \& University of Maryland, Baltimore County, MD, USA}

\begin{abstract}
A cornerstone of special relativity is Lorentz Invariance, the postulate that all observers measure exactly the same photon speeds independently on the photon energies. However, a hypothesized structure of spacetime may alter this conclusion at ultra-small length scales, a possibility allowed in many of the Quantum-Gravity (QG) formalisms currently investigated. A generalized uncertainty principle suggests that such effects might occur for photon energies approaching the Planck energy, $E_{Planck}=M_{Planck} c^2 \simeq 1.22\times10^{19} GeV$. Even though all photons yet detected have energies $E_{ph}<<E_{Planck}$, even a tiny variation in the speed of light, when accumulated over cosmological light-travel times, may be revealed by high temporal-resolution measurements of sharp features in Gamma-Ray Burst (GRB) lightcurves. Here we report the results of a study using the emission from GRB~090510 as detected by \textit{Fermi}'s LAT and GBM instruments, that set unprecedented limits on the dependence of the speed of photons on their energy. We find that the mass/energy scale for a linear in energy dispersion must be well above the Planck scale, something that renders any affected QG models highly implausible.
\end{abstract}

\maketitle

\thispagestyle{fancy}

\section{Introduction}
Some Quantum-Gravity models (see \cite{liv1,liv2,liv3,liv4,liv5} and references therein) postulate an inherent structure of spacetime (e.g. either a foamy, a discrete, or a lumpy spacetime) near the Planck scale $\lambda_{Pl}=\sqrt{G\hbar /c^3} \sim 10^{-35}m$. Since Lorentz symmetry is scale invariant (i.e. all scales are equivalent), the postulated existence of a special scale due to such QG effects can possibly lead to violations of Lorentz Invariance (LIV). One manifestation of such violations could be a dispersion in photon propagation, in which the speed of a photon in vacuo becomes dependent on its energy. 

A detection and measurement of LIV effects would be indisputably invaluable to our understanding of the nature of spacetime at extremely small scales. Furthermore, even setting an upper limit on the magnitude of such effects can still prove very valuable, since measurements affected by physics at tiny Planck scales are rare and hard to perform, and can still be used to constrain and guide the relevant research. In this case, we have used high-quality measurements from GRB~090510 performed with both the GBM and LAT instruments on board the \textit{Fermi} observatory to set the most stringent limits to date on the magnitude of such energy-dispersion effects. 

In the following, a brief introduction to the relevant formalism of LIV will be given, and some of the properties of GRB 090510 will be described. For more information on this study please refer to our Nature publication and its associated supplementary information \cite{nature}.

\subsubsection*{Lorentz-Invariance Violation}
Consider two photons of energies $E_h>E_l$ emitted simultaneously from a distant astrophysical source at redshift $z$. According to the postulated LIV effects these two photons will travel with different velocities and will arrive with a time delay $\Delta t$ equal to:

\begin{equation}
\label{eqDT}
$$ $\Delta t=s_{n}\frac{(1+n)}{2H_{0}}\frac{(E_{h}^{n}-E_{l}^{n})}{(M_{QG,n}c^{2})^{n}}\int_{0}^{z}\frac{(1+z')^{n}}{\sqrt{\Omega_{m}(1+z')^{3}+\Omega_{\Lambda}}}dz'$ $$
\end{equation}

Here $M_{QG}$ is the ``Quantum-Gravity mass'', a parameter that sets the energy scale at which such QG effects start to become important. Its value is assumed to be near the Planck Mass ($M_{Pl}=\hbar c/ \lambda_{Pl} \sim 10^{19}$GeV/$c^2$) and most likely smaller than it. The model-dependent parameter $n$ sets the order of the LIV and is assumed to be one or two, corresponding to linear ($\Delta t\propto \Delta E/M_{QG}$, with $\Delta E\equiv E_{h}-E_{l} \simeq E_{h}$) and quadratic ($\Delta t\propto \left(E_{h}/M_{QG}\right)^2$) LIV respectively. The model-dependent parameter $s_n$ is equal to plus or minus one, and it sets the type of LIV: a positive (negative) $s_n$ corresponds to a positive (negative) time delay or a speed retardation (acceleration) with an increasing photon energy. 

In this analysis we used the times and the energies of the detected photons from GRB 090510 to set an upper limit on the strength of any LIV effects or equivalently a lower limit on $M_{QG}$. Since $M_{QG}$ is expected to be close and most likely smaller than the Planck Mass, a limit that is close to or, even better, over the Planck Mass is especially physically meaningful since it excludes most of the allowed parameter space, rendering any affected models highly implausible. It should be noted that not all QG models that predict a spacetime structure actually require such an energy dispersion. Instead, many of the models are consistent with LIV but they do not explicitly require it. A model that actually requires LIV, and can be directly constrained by such limits, is the stringy-foam model described in \cite{el08}.

\subsubsection*{GRB 090510}
Because of their short duration (typically with short substructure in the form of a series of narrow spikes) and cosmological distances, GRBs are well-suited for constraining LIV. In this study, we used measurements on the bright and short GRB 090510, which triggered both the LAT and GBM on May 10th 2009 at 00:22:59.97 UT (hereafter all times are measured relative to this trigger time). Ground-based optical follow-up spectroscopic data \cite{Rau2009}, taken 3.5 days later, exhibited prominent emission lines at a common redshift of $z=0.903\pm0.003$. The GBM light curve (figures \ref{fig:ET} b,c; 8 keV -- 40MeV) consisted of 7 main pulses. After the first dim short spike near trigger-time (hereafter called the ``precursor''), the flux went down to background levels. The main emission as detected by the GBM started at 0.53s and lasted for $\lesssim$0.5s. On the other hand, the main emission as detected by the LAT (figures \ref{fig:ET} a, d--f; E$\gtrsim$20MeV) started after the main GBM emission (0.65s vs 0.52s) and also extended to a significantly longer time scale (than the GBM emission) of about 200s. The emission detected by the LAT extended to an energy of about 31GeV, which is the highest energy ever detected from a short GRB. The fact that this 31GeV photon was detected shortly after the beginning of the burst ($\sim$0.8s), and the fact that the LAT-detected emission exhibited a series of very narrow spikes (up to few tens of ms width) that extended to high ($>$ tens of MeV) energies, allowed us to set stringent limits on the Quantum-Gravity mass. We used two independent methods to set these limits, described in the following section.

\section{Lower Limits on the Quantum-Gravity Mass}
\subsection{Using the arrival time of a single photon}
According to equation \ref{eqDT}, if we knew the LIV-induced time delay (${\Delta}t$) between two photons of energies $E_h>E_l$, then we should be able to make a measurement on $M_{QG}$ (for some model dependent parameters $s_n$ and $n$, and for a known measured $z$). However, this time delay cannot be measured because we do not know the exact emission times and because we cannot safely assume that the two photons were emitted from the same location (hence assume that there were no extra propagation delays due to non co-location). However, what we can do is to first assume that the higher-energy (HE) photon was emitted some time during a lower-energy (LE) emission episode (which starts from time $t_{start}$ and extends up to at least the HE-photon detection time $t_{HE}$), then calculate a \textit{maximum} time delay for the HE photon as ${\Delta}t_{max} = t_{HE}-t_{start}$, and finally calculate using equation \ref{eqDT} a \textit{lower limit} on the Quantum-Gravity mass. 

Since the end time of the lower-energy emission episode cannot be safely assumed, the method mentioned above only works to constrain \textit{positive} time delays ($s_n=+1$). Also, this method is still valid even if the HE photon and the majority of the photons comprising the associated LE-emission episode are not emitted from the same exact location. Specifically, it only requires that there should be at least one LE photon that was emitted from the same location as the HE photon and was detected during the lower-energy emission episode, something that is generally safe to assume. Lastly, this method is only very weakly sensitive to any possible spectral lags occurring intrinsically at the GRB. Such lags have not been observed in short GRBs: they have been observed only in sub-MeV energies (and not in the higher-energy range of interest here), and even when they are observed (i.e. in long GRBs), they are of the order of a typical spike width (few tens of ms), which is considerably smaller than the time delays typically used in this analysis. As it will be shown, spectral lags, if they are actually present in GRB~090510, would have a negligible effect on our main results and would certainly not affect the conclusion of this study. 

Setting stringent limits on the Quantum-Gravity mass requires an as-high-as-possible $E_h$ ($E_h$ usually ${\gg}E_l$ so $E_h-E_l{\simeq}E_h$) and an-as small-as-possible ${\Delta}t$. For GRB 090510, we used the highest-energy photon detected (31GeV) for settings such limits. Even if another lower-energy photon corresponded to a more stringent limit, to be conservative, we only used the 31GeV photon since its larger time delay makes it less sensitive to uncertainties in the choice of $t_{start}$ and to any intrinsic spectral lags. Furthermore, its higher energy renders it less probable of being a background event than an alternative lower-energy candidate. In the following calculations, again to be conservative, we used values for the GRB's redshift and for the 31GeV photon's energy reduced by one standard deviation. 

The 31GeV photon had a reconstructed energy of $30.53^{36.32}_{27.97}$GeV ($1\sigma$ confidence intervals), it was detected 0.829s post-trigger, and it coincided with one of the GBM pulses. Very thorough analyses confirmed this event to be a real photon (instead of a background cosmic ray) associated with this GRB (e.g. from the diffuse galactic or extra-galactic emission). The properties of its associated event (absence of energy deposition in the anti-coincidence detector, the signature of an electromagnetic shower in the instrument, etc.) and the results from our event classification algorithms strongly supported its gamma-ray nature. Furthermore, its strong temporal (detected during the prompt emission) and directional (less than one PSF from the reconstructed by Swift direction) coincidence with the GRB supported its association with this source. And independently of these considerations, simple analysis regarding the rate of expected background events (both cosmic- and gamma-rays) resulted in a probability of less than $\sim10^{-7}$ ($>5\sigma$) of the background producing such an event.

\begin{figure*}[t]
\label{fig:ET}
\centering
\includegraphics[width=135mm]{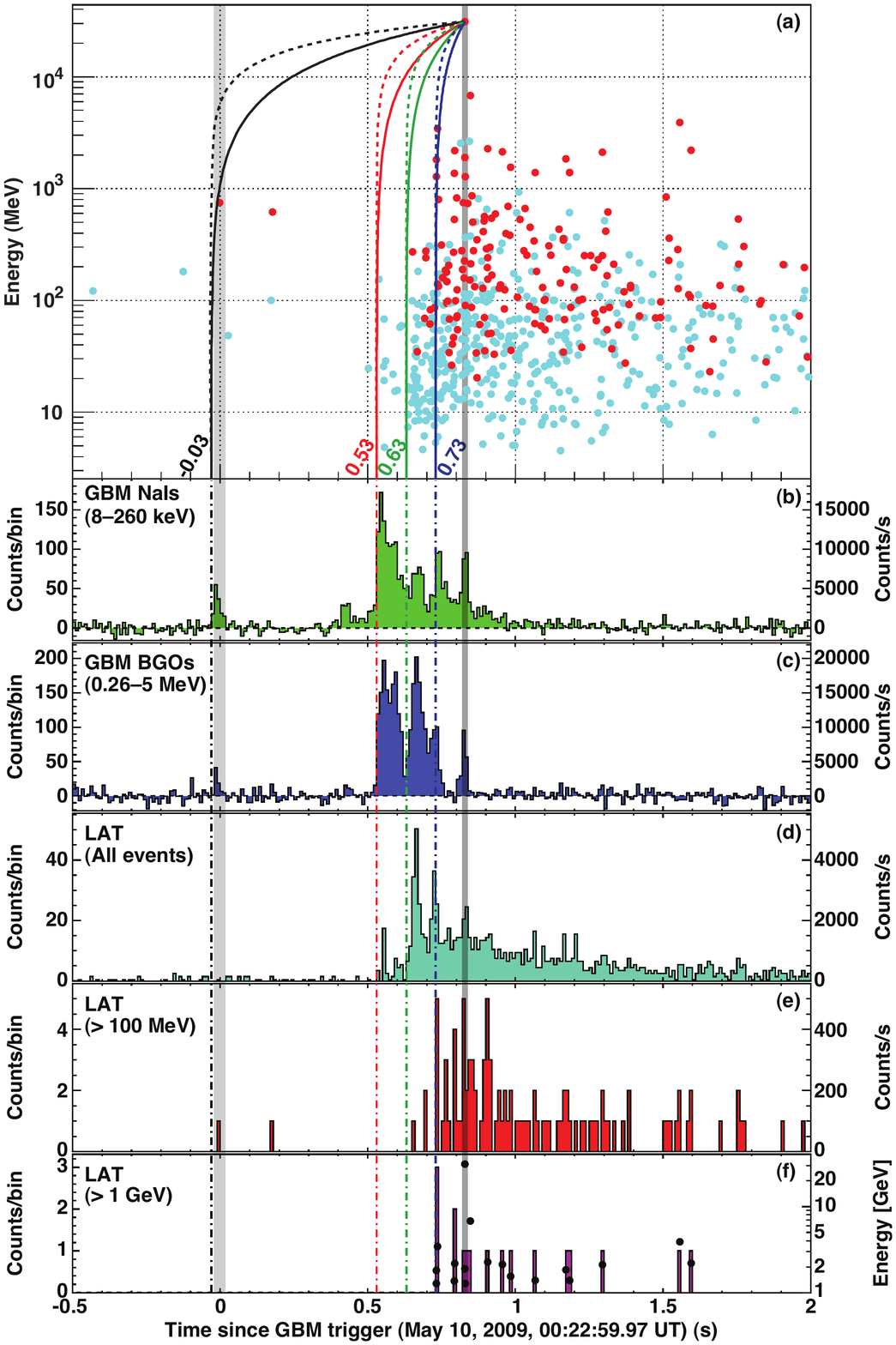}
\caption{Panel (a): energy vs. arrival time with respect to the GBM trigger time for the LAT photons that passed the transient event
selection (red) and the photons that passed the onboard $\gamma$-ray filter (cyan). The solid 
and dashed curves are normalized to pass through the 31GeV photon and represent the relation between a photon's energy and arrival time 
for linear (n=1) and quadratic (n=2) LIV, respectively, assuming it is emitted at $t_{start}=-30ms$ (black; first small GBM pulse onset), 530ms (red; 
main $<MeV$ emission onset), 648ms (green; $>100MeV$ emission onset), 730ms (blue; >GeV emission onset). Photons emitted at $t_{start}$ would be located 
along such a line due to LIV time delays. Panels (b)--(f): GBM and LAT lightcurves, from lowest to highest energies. Panel (f) also 
overlays energy vs. arrival time for each photon, with the energy scale displayed on the right side. The dashed-dotted vertical lines show our 4
different possible choices for $t_{start}$. The gray shaded regions indicate the arrival time of the 31GeV photon (on the right) and of a 
750MeV photon (during the first GBM pulse) (on the left), which can both constrain a negative time delay. Panels (b) and (c) show background 
subtracted lightcurves for the GBM detectors. Panels (d)--(f) show, respectively, LAT events passing the onboard $\gamma$-ray filter, LAT transient class events with $E>100MeV$, and LAT transient class events with $E>1$ GeV. In all lightcurves, the time-bin width is 10ms.
} 
\end{figure*}

The most conservative and of very high confidence assumption that can be made regarding the possible emission time of the 31GeV photon was that \textit{it was not emitted before the beginning of the precursor} (30ms before the trigger). For such an assumption, ${\Delta}t=0.829+0.03=0.859s$ and the associated lower limit on the QG mass for linear energy dispersion ($n=1$) is $M_{QG,1}\gtrsim1.19{\times}M_{Planck}$. An illustration that corresponds to this choice of $t_{start}$ is shown in sub-figure \ref{fig:ET}a with the black solid (n=1) and dashed (n=2) lines. It should be noted that the validity of this assumption depends on the absence of a second undetected precursor before the detected one. Even though the \textit{Fermi} LAT and GBM did not show evidence of any such second precursor, and even though no such two-precursor short GRBs have been detected, the existence of a second precursor starting at time $t_{pre,2}$ before the start of the first detected precursor $t_{pre,1}=-0.03s$ would decrease this most conservative limit by a factor of $\frac{t_{HE}-t_{pre,2}}{t_{HE}-t_{start}}$. The presence of any intrinsic spectral lags would correspond to a similar change in the limits. A negative spectral lag of $t_{lag} \sim 0.01s$ (in which the 31 GeV photon was emitted $t_{lag}$ time before the lower-energy associated events) would correspond to a decrease of our lower limit by a factor of $\frac{\Delta t + t_{lag}}{\Delta t} \simeq 1$ since here $t_{lag}<<\Delta t$. 

Our most conservative limit, described above, can be improved by considering that it is actually very likely that the 31GeV photon is associated with only the main emission starting at $t_{start}=0.53s$ instead of with the whole burst (i.e. assumming that the 31GeV photon is far more likely to have been emitted during the main emission episode instead someting before it). This assumption is strongly supported by the fact that the 31GeV photon is expected to be emitted together with other lower-energy (sub MeV or MeV) photons that do not suffer from LIV delays and therefore mark its emission time. Using this assumption for the relevant emission time, the maximum time delay for the 31GeV photon becomes considerably smaller and our limits significantly stronger $M_{QG,1}{\gtrsim}3.42{\times}M_{Planck}$. Similarly, we can make somewhat less conservative assumptions regarding the lower-energy emission interval and associate the 31GeV photon with the start of the $>$100MeV emission or the start of the $>$1GeV emission, yielding stronger  limits, yet with less confidence. Such limits correspond to smaller time delays, therefore they are relatively more sensitive to uncertainties in the choice of $t_{start}$ (because of now having to choose $t_{start}$ based on fewer photons) and to any possible intrinsic spectral lags. The results from all possible associations are shown in table I. As above, these limits are illustrated in sub-figure \ref{fig:ET}a. It should be noted that all the results of this method yield strong limits on $M_{QG}$ for linear LIV $n=1$, while they give considerably less constraining limits for the quadratic case. Also, as mentioned before, the above results are for the case of speed retardation $s_{n}=+1$ since we cannot safely assume the \textit{latest} emission time the 31GeV photon was emitted.

Finally, we note that the 31GeV photon arrives near the peak of a very bright and narrow spike in the soft gamma-ray lightcurve, which has a width of $\sim$10--20 ms (see figure \ref{fig:ET}). Such an association is not secure since the 31GeV photon could have landed over the spike just by chance. 
However, just as an idea regarding the magnitude of the limits that could be attainable provided such an association was secure, we 
calculate a significantly higher limit of $M_{QG,1}\gtrsim102{\times} M_{Planck}$ for both speed retardation and acceleration $s_{n}=\pm1$ (since now we can constrain both the earliest and latest possible emission time of the 31GeV photon). Similarly, a weaker 
but independent and somewhat more robust limit on a possible negative time delay (which constrains the super-luminal case, $s_{n}=-1$) may be 
obtained from the $\sim$0.75GeV photon that is observed during the precursor. This photon has a high probability 
of being from GRB 090510 (a chance probability of the background producing such a photon of $\sim$1.2$\times 10^{-6}$ corresponding to $\sim 4.6 \sigma$), and the 1$\sigma$ confidence interval 
for its energy is 693.6--854.4MeV. For the case of speed acceleration ($s_{n}=-1$), we can calculate an \textit{maximum} time delay of ${\Delta}t$<19ms and a limit of $M_{QG,1}\gtrsim1.33\times M_{Planck}$ (using again a reduced-by-one-standard-deviation photon energy in the calculation). These two associations with an individual spike are shown in figure \ref{fig:ET}a with the vertical dashed bands and in table I.

\subsection{Using the DisCan method}

An alternative method was also used for constraining any linear in energy LIV effects. This method, called DisCan \cite{DisCan} (Dispersion Cancellation), extracts dispersion information from \textit{all the LAT-detected photons} ($\sim$30 MeV -- $\sim$30 GeV) and \textit{does not involve binning in time or energy}. It is based on the fact that any QG-induced time delays would be expected to smear the spiky structure of the lightcurve. The DisCan method applies different trial spectral lags (time delays that are inversely proportional to the energy) to the lightcurve until it finds the one that maximizes a measure of its ``sharpness''. The trial spectral lag that accomplishes this is equal and opposite in sign to the sum of any QG-induced and intrinsic-to-the-GRB spectral lags. 

\begin{figure}[t]
\centering
\includegraphics[width=80mm]{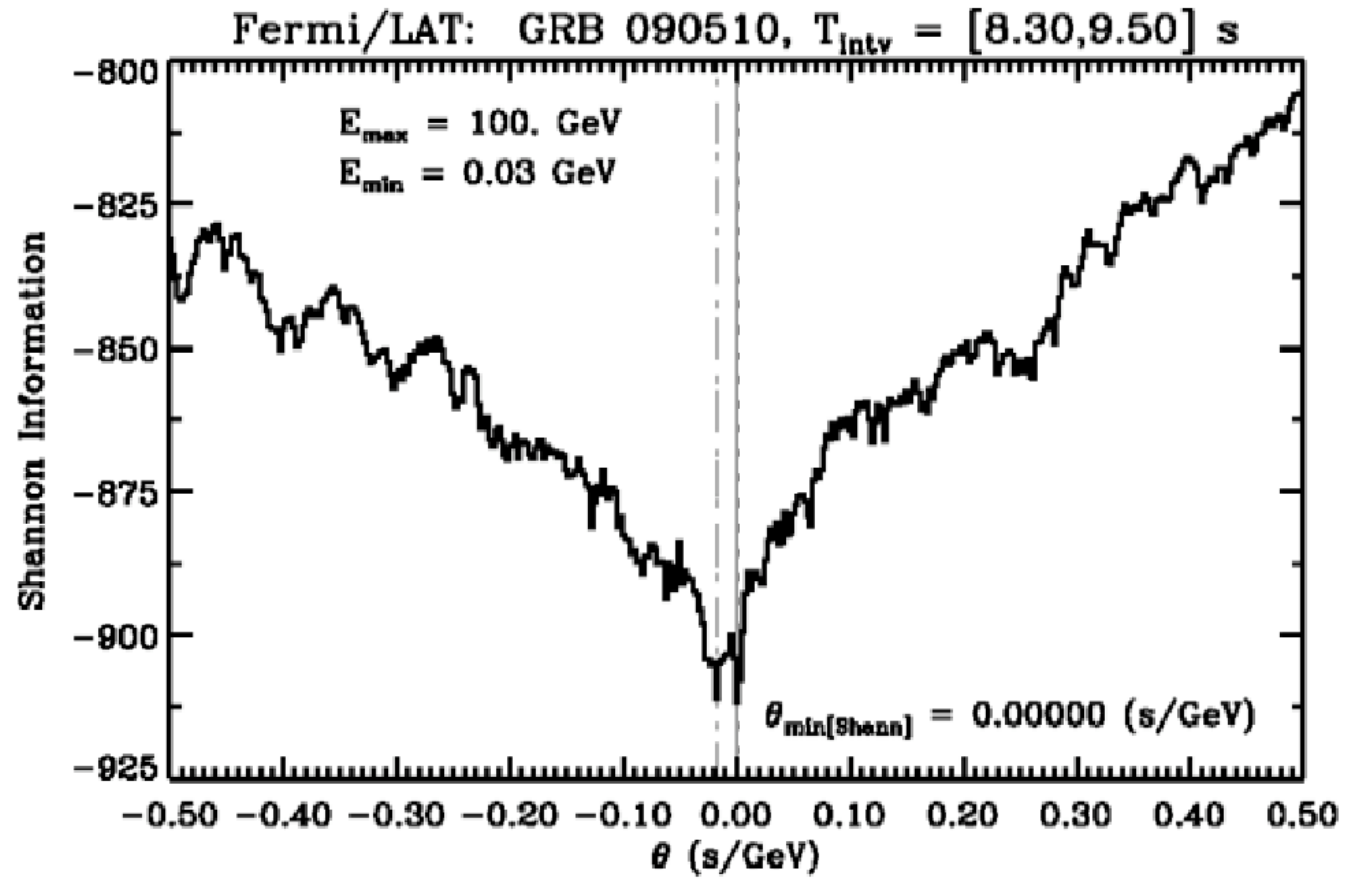}
\caption{The value of the cost function (Shannon Information) versus the trial spectral lag value ($\theta$). The best
value of $\theta$, equal to zero, is annotated, and shown as a vertical solid line. The two dashed vertical lines left and right
of the best value represent the $\theta$ values which are 100 times less probable (-18 and +2 ms/GeV respectively) than the best $\theta$ value, \textit{for the given data set}. Thus the contained interval between the two dashed lines is an approximate error region,
but does not reflect statistical uncertainties.}
\label{fig:DisCan1}
\end{figure}

The DisCan method was applied using the photons detected by the LAT during the interval 0.5-1.5s post-trigger, the burst interval with the most intense emission. The results were not significantly sensitive to any variations on the stop times for the interval ($\pm$0.25s) or the energy upper limits (1, 3 \& 100 GeV). The sharpness of the lightcurve was measured by using a cost function: the ``Shannon Information'' (eq. 11 of \cite{DisCan}). The value of this cost function is equal to the entropy (modulo a minus sign), and the difference of two values of the cost function are equal to the relative probability that one value is more likely for a given data set. Figure \ref{fig:DisCan1} shows the value of this cost function versus the trial spectral lag value. The minimum value of the cost function (most probable value) is $0^{+2}_{-18}$ ms/GeV. The errors here correspond to the trial spectral lag values that are 100 times less probable than the best value of 0ms/GeV, and are shown with the two vertical dashed lines in figure \ref{fig:DisCan1}.

\begin{figure}[t]
\centering
\includegraphics[width=80mm]{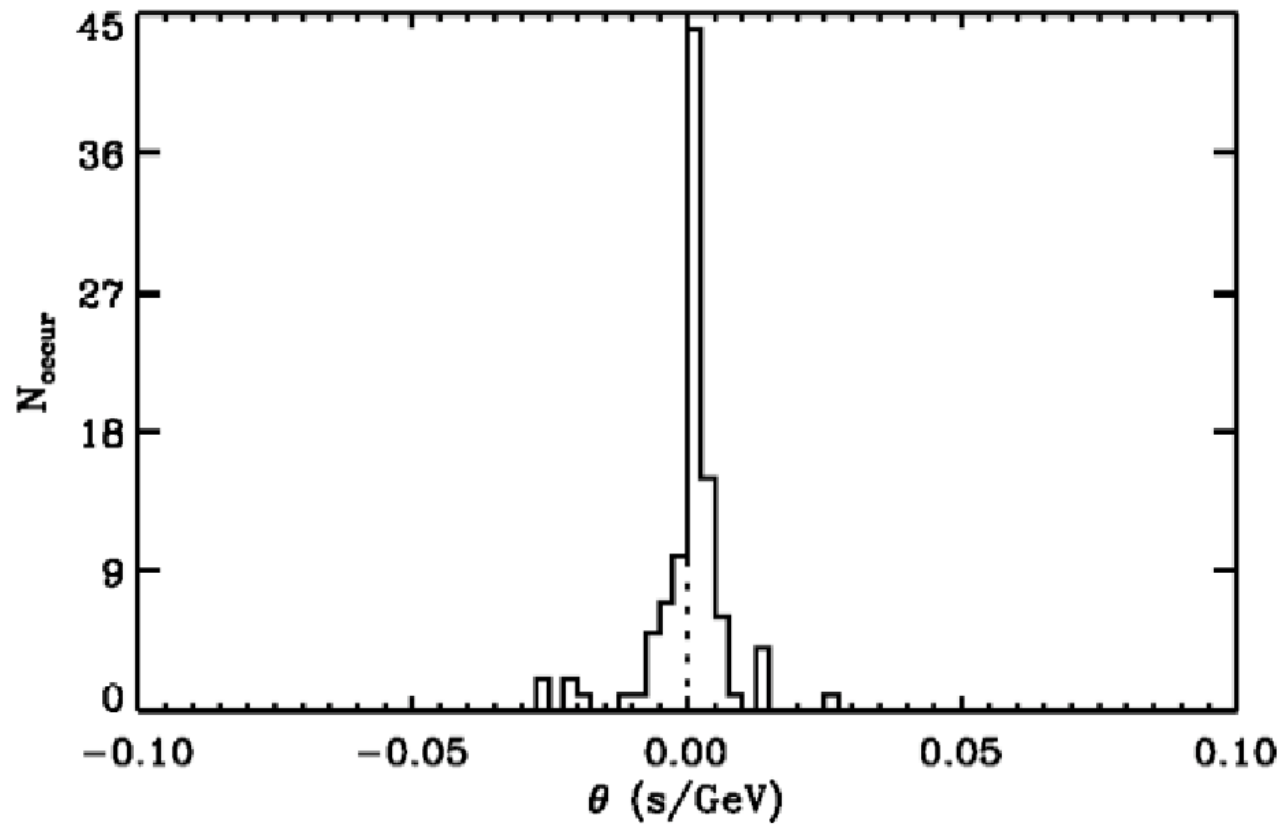}
\caption{For the interval analyzed in Figure \ref{fig:DisCan1}, to gauge uncertainty due to statistical variations
we generated 100 realizations with the photon times randomized. This figure shows the distribution of the minimum trial spectral-lag value $\theta_{min}$ for these 100 realizations. From this distribution, the 99\%CL for $\theta_{min}$ is 30ms/GeV.}
\label{fig:DisCan2}
\end{figure}

However, this result does not comprise a definite measurement of the spectral lag value; instead it is just one of the results that are compatible with the inherent uncertainties associated with our choice of time interval and energy range and with the limited statistics of the dataset. To estimate these uncertainties, a bootstrap analysis was performed, in which the DisCan method was applied to a randomized data set (a set produced by randomizing the association between the energies and times of the events). The reassignment destroyed any correlation with the energy, and therefore it removed any spectral lags. As a result, the application of this method to a randomized dataset should on average measure a spectral lag value that is equal to zero, while the width of the distribution of spectral lag values would be a measure of the statistical uncertainty of the measurement associated with our dataset. The results of the bootstrap analysis are shown in figure \ref{fig:DisCan2}, which shows the distribution of the measured spectral lags for each of the 100 randomized datasets. The measured spectral lags lie within a value of $<30ms/GeV$ for 99\% of the cases, and within a value of $<10ms/GeV$ for 90\% of the cases.

The combined result from minimizing the cost function for the actual GRB 090510 dataset and from the bootstrap analysis on the randomized datasets is an upper limit on the energy dispersion equal to $<$30 ms/GeV or $M_{QG}>1.22 \times M_{Planck}$ for linear energy dispersion of either sign $s_n=\pm 1$ at the 99\% confidence level. This result is shown in table I, along with the results from the previous method. For reasons similar to those advanced in the previous subsection (improbability of intrinsic lags or fortuitous cancellation of quantum gravity and intrinsic dispersion) we take this result as an upper limit on LIV-induced dispersion (i.e. we ignore any possible intrinsic spectral lags).

\begin{table*}[ht]
\label{table}
\begin{center}
\small{
\begin{tabular}{|c|c|c|c|c|c|c|c|c|c|}
\hline
& $t_{start}$ & Limit on $|\Delta t|$ & Reasoning for $t_{start}$ or method& $E_{l}$ & Valid for $s_{n}$ & Confidence& Limit on $M_{QG,1}$ & Limit on $M_{QG,2}$\tabularnewline
 & (ms) & (ms) &used for setting the limits & (MeV) &  &  & ($M_{Planck}$) & ($10^{10}GeV/c^{2}$)\tabularnewline
\hline
\hline 
(a) & -30 & < 859 & start of any $<$MeV emission & 0.1 & +1 & very high & > 1.19 & > 2.99\tabularnewline
\hline 
(b) & 530 & < 299 & start of main$<$MeV emission & 0.1 & +1 & high & > 3.42 & >5.06\tabularnewline
\hline 
(c) & 630 & < 199 & start of main$>$0.1 GeV emission & 100 & +1 & high & > 5.12 & > 6.20\tabularnewline
\hline 
(d) & 730 & < 99 & start of main$>$1 GeV emission & 1000 & +1 & medium & > 10.0 & > 8.79\tabularnewline
\hline 
(e) & -- & < 10 & association with$<$1 MeV spike & 0.1 & $\pm1$ & low & > 102 & > 27.7 \tabularnewline
\hline 
(f) & -- & < 19 & if 0.75GeV $\gamma$-ray from 1st spike&  & -1 & low & > 1.33 & > 0.54 \tabularnewline
\hline 
(g) & \multicolumn{2}{c|}{$|\Delta t/\Delta E|<30ms/GeV$} & Lag analysis of all LAT photons& -- & $\pm1$ & very high & > 1.22 & --\tabularnewline
\hline
\end{tabular}
}
\caption{Lower limits on the Quantum-Gravity mass scale associated with possible Lorentz
Invariance Violation, that we can place from the lack of time delay in the arrival of high-energy
photons relative to low energy photons, from our observations of GRB 090510.}
\end{center}
\end{table*}

\section{Conclusion}
We have used high-quality measurements on GRB 090510 performed by the GBM and LAT instruments on board the \textit{Fermi} spacecraft to constrain tiny variations on the speed of light in vacuo that are linear or quadratic to its energy. We used two independent methods to obtain conservative and unprecedented upper limits on the magnitude of such speed variations. Our limits ($M_{QG,1}\gtrsim$few$\times M_{Planck}$) strongly disfavour any models predicting such linear-in-energy variations in the speed of light. Limits of such strength (over the Planck mass) are especially physically meaningful, since they exclude almost all of the allowed parameter space, rendering any affected models highly implausible. Any models predicting quadratic LIV are not significantly affected by our results. Our limits can be used to guide future research in Quantum Gravity and in general give insight on the nature of spacetime at minuscule Planck scales.

\bigskip % extra skip inserted

\begin{acknowledgments}
The Fermi LAT Collaboration acknowledges support from a number of agencies and institutes
 for both the development and the operation of the LAT as well as scientific data analysis.
 These include NASA and DOE in the United States, CEA/Irfu and IN2P3/CNRS in France, ASI and
 INFN in Italy, MEXT, KEK, and JAXA in Japan, and the K. A. Wallenberg Foundation, the Swedish
 Research Council and the National Space Board in Sweden. Additional support from INAF in Italy
 for science analysis during the operations phase is also gratefully acknowledged. 
The Fermi GBM Collaboration acknowledges the support of NASA in the United States and DRL
 in Germany.
\end{acknowledgments}

\bigskip % extra skip inserted

\end{document}